\documentstyle[a4wide,12pt,epsf]{article}
\begin{document}
\begin{flushright}
\baselineskip=12pt
{SUSX-TH-98-021}\\
{hep-ph/9812283}\\
{November 1998}
\end{flushright}

\begin{center}
{\LARGE \bf M-THEORY DARK MATTER \\}
\vglue 0.35cm
{D.BAILIN$^{\clubsuit}$ \footnote
{D.Bailin@sussex.ac.uk}, G. V. KRANIOTIS$^{\spadesuit}$ \footnote
 {G.Kraniotis@rhbnc.ac.uk} and A. LOVE$^{\spadesuit}$ \\}
	{$\clubsuit$ \it  Centre for Theoretical Physics, \\}
{\it University of Sussex,\\}
{\it Brighton BN1 9QJ, U.K. \\}
{$\spadesuit$ \it  Department of Physics, \\}
{\it Royal Holloway and Bedford New College, \\}
{\it  University of London,Egham, \\}
{\it Surrey TW20-0EX, U.K. \\}
\baselineskip=12pt

\vglue 0.25cm
ABSTRACT
\end{center}

{\rightskip=3pc
\leftskip=3pc
\noindent
\baselineskip=20pt
The phenomenological implications of the eleven dimensional limit of 
$M$-theory (strongly coupled $E_8\times E_8$) are investigated. In particular 
we calculate the supersymmetric particle 
spectrum subject to constraints of correct 
electroweak symmetry breaking and the requirement that the lightest 
supersymmetric particle provides the dark matter of the universe.
We also calculate direct detection event rates of the lightest neutralino
relevant for non-baryonic dark matter experiments. The modulation effect,
due to Earth's annual motion is also calculated.}

\vfill\eject
\setcounter{page}{1}
\pagestyle{plain}
\baselineskip=14pt

\section{Introduction}
In recent years it has become clear that the five perturbative 
string theories and the 11 dimensional supergravity are different 
limits in moduli space of a unique fundamental theory. This 
strongly indicates an  enormous degree of symmetry of the underlying 
theory and its intrinsically non-perturbative nature. 
String duality correlates the six corners of the moduli space. 
The duality transformation involves Planck's constant $\hbar$ 
and is therefore  essentially quantum mechanical. 
One then might argue that before we have the complete picture of 
$M$(other)theory 
it is premature to make any attempt at phenomenology. 
However, it may be that the corners 
of the moduli space capture most of the features of the 
theory relevant for low-energy phenomenology \footnote{The 11-D limit 
of  $M$-theory is not gauge invariant so Horava and Witten have argued that 
quantum terms are needed to restore gauge invariance \cite{HORWIT}.
See however, M. Faux's 
argument for a consistent classical limit of $M$-theory \cite{FAUX}}.

One of the most interesting dualities 
(and the most relevant for low-energy phenomenology) 
is the one in which the low-energy limit of 
$M$-theory (i.e 11D-Supergravity) compactified on the line segment 
$I\sim S^1/\bf{Z}_2$ (i.e an orbifold) is equivalent to the 
strong coupling limit of the $E_8\times E_8$ heterotic string 
\cite{HORWIT}. 
In this picture on one end of the line segment 
of length $\pi \rho$ live the 
observable gauge fields contained in 
the first $E_8$ while the hidden sector fields live in
the second $E_8$ factor
on the other end. Gravitational fields propagate in the bulk.

The main phenomenological virtue of such a framework is that due to 
the extra dimension one may obtain unification of {\it all} interactions 
at a scale $M_U\sim 3\times 10^{16} GeV$ consistent with experimental data 
for the low energy gauge couplings \cite{witten}. 
Also the analysis of gaugino condensation reveals that phenomenologically 
acceptable gaugino masses, 
comparable with the gravitino mass $m_{3/2}$, 
arise quite naturally in sharp contrast to  
the weakly coupled case where tiny gaugino masses were troublesome 
\cite{NILLES}.
It is therefore of great importance to investigate further the 
phenomenological implications of the 11-dimensional low energy limit 
of $M$-theory and to determine any deviations from the weakly coupled 
case.

It is well known from observation (rotation curves) 
and supported by theoretical 
reasons (inflation) that most of the matter in our galaxy, and in the 
universe in general consists of a new exotic form of matter beyond the  
standard baryonic matter. The  identification of this new 
form of matter 
is one of the most important challenges  of astroparticle physics. 
A convenient parametrization of matter in the universe is given by 
the quantity, $\Omega=\rho/\rho_c$, where $\rho$ is the matter 
density of the universe  $\rho_c=3 H^2/8\pi G_N$ is the critical 
matter needed to close the universe, and $H$ is the Hubble parameter 
which is generally parametrized by $H=100 h$ Km/s Mpc. Experimentally 
the current evaluations of $h$ lie in the interval 
$0.4\leq h \leq 0.8$. Inflation predicts generically $\Omega=1$, and
the quantity that enters in the particle physics analysis is 
$\Omega h^2$.  The COBE data indicates that there is more 
than one component to the non-baryonic dark matter, namely , a 
hot component which consists of particles which would be relativistic 
at the time of galaxy formation and a cold component which would 
be non-relativistic at the time of galaxy formation. The hot 
component could be massive neutrinos and the cold component could be 
either axions, or supersymmetric  particles.
In supersymmetric  theories 
with $R$-parity invariance 
the lightest supersymmetric particle (LSP) is 
stable and a very good candidate for cold dark matter
\cite{EHNOS}. For most regions 
of the parameter space the LSP is the lightest neutralino.
The axion mass relevant for cosmology  has been constrained 
recently by experiment \cite{DAWN}. On the other hand 
experiments for the direct detection of the neutralino have reported 
progress and soon their sensitivity will start exploring the 
susy parameter space \footnote{In fact recently the DAMA/NaI Collaboration
reported an indication of a possible modulation effect in  
direct detection experiments for neutralinos \cite{BERNABEI}.}.
It is therefore of great importance  to calculate the cross section 
of neutralinos with detector nuclei in the $M$-theory framework since 
such searches are complementary to accelerator experiments.
In the analysis of this paper we shall assume the parameter space 
of the effective supergravity from $M$-theory is limited by
\begin{equation}
0.1\leq \Omega_{LSP} h^2 \leq 0.4
\end{equation} 
where the LSP is identified with the lightest neutralino
\footnote{In string theory one can entertain the 
possibility of superheavy dark matter. For a recent work 
see Benakli et al \cite{KARIM}.}. We will discuss 
later on the effect of varying the allowed range of $\Omega_{LSP} h^2$.

Several 
papers have recently  analyzed  
the effective supergravity and the soft supersymmetry-breaking terms 
emerging in the $M$-theory framework
\cite{ANTO,NANO,NILLES,DUDAS,STEVE,
CHOI,LUKAS,LI}. Some properties of the sparticle spectrum 
which depend only on the boundary conditions and not on the details of 
the electroweak symmetry breaking, have been discussed in \cite{CHOI}.
Further phenomenological implications for the supersymmetric 
particle spectrum 
including the constraint of correct electroweak 
symmetry breaking 
\cite{Tam:Rad}, have been performed in \cite{BKL,KINE,CASA}. 
It is the purpose of this paper to 
investigate the phenomenological implications of $M$-theory relevant 
to accelerator experiments and the cosmological properties of the LSP, 
as well as the 
prospects for its detection in 
underground non-baryonic dark matter experiments.
Some of our findings were already reported briefly elsewhere \cite{BKL}.
Here we shall present our calculations in greater detail, and their 
connection with experiments to detect the LSP as well as presenting 
further results for additional values of the goldstino angle.

\section{Soft supersymmetry breaking-terms in $M$-theory}
The soft supersymmetry-breaking 
terms are determined by the following functions of the effective 
supergravity theory \cite{LUKAS,CHOI}:
\begin{eqnarray}
K&=&-{\rm ln}(S+\bar{S})-3{\rm ln}(T+\bar{T})+\Biggl(\frac{3}{T+\bar{T}}+
\frac{\alpha}{S+\bar{S}}\Biggr)|C|^2, \nonumber \\
f_{E_6}&=&S+\alpha T, \;\;f_{E_8}=S-\alpha T, \nonumber \\
W&=&d_{pqr}C^p C^q C^r
\label{mfunc}
\end{eqnarray}
where $K$ is the K$\rm{\ddot{a}}$hler potential, $W$ the 
perturbative superpotential, and 
$f_{E_6}, f_{E_8}$ are the gauge kinetic functions for the observable 
and hidden sector gauge groups $E_6$ and $E_8$ respectively. 
Also $S,T$ are the dilaton and 
Calabi-Yau moduli fields and $C^{\alpha}$ charged matter fields.
The superpotential
and the gauge kinetic functions are exact up to non-perturbative effects. 
The integer $\alpha=\frac{1}{8\pi^2}\int J \wedge 
[{\rm tr}(F\wedge F)-\frac{1}{2}{\rm tr}(R\wedge R)]$ for the 
K$\rm{\ddot a}$hler form 
$J$ is normalized as the generator of the integer (1,1) cohomology.
Let us give an example of how one can calculate this important 
parameter in Calabi-Yau compactifications of $M$-theory
\footnote{For other examples see \cite{BEN,STIEB}.}.
For instance in the $(0,2)$ Calabi Yau models the following 
relations in the Chern classes hold:
\begin{eqnarray}
c_2({\cal M})&=&c_2(V_1)+c_2(V_2) \nonumber \\
c_1(V_{1,2})&=&0\;\rm{mod} 2
\end{eqnarray}
We now calculate 
$\alpha$, in Calabi-Yau manifolds defined as intersections 
in products of complex projective spaces $P^{n}_{\omega_:\cdots:
\omega_n}$. Then a polynomial $f_a(z)$ in the quasi-homogeneous 
coordinates of $P^{n}_{\bf{W}}$ of total weight $q_a$ is a section 
of the line bundle $O(P^n_{\bf{W}})$ and the Chern classes expand 
as 
\begin{equation}
c[{\cal M}||{\cal E}]=\frac{\prod_{r=1}^m\prod_{i=0}^{n_r}(1+\omega_i^{(r)}
J_r)}{\prod_{a=1}^{k}(1+\sum_{s=1}^{m}q_a^sJ_s)}
\end{equation}
For example  in the model of Kachru \cite{KAC} in which ${\cal M}$ is 
defined as a complete 
intersection of two degree six polynomials $P_1$ and $P_2$ in 
$P_{1,1,2,2,3,3}^{5}$ if we choose the instanston numbers as
$V_1=(3,3,2,;1,1,1,1,1,1,1,1)$ and $V_2=(6,3;1,1,1,2,2,2)$
then $c_2(V_1)=7$ and $c_2(V_2)=15$ while the Chern class of 
the tangent bundle is $c_2({\cal M})=22$. 
In that case
\begin{eqnarray}
\alpha_{obs}&=&(c_2(V_1)-\frac{1}{2}c_2({\cal M}))\int J \wedge J \wedge J
=-4 \nonumber \\
\alpha_{hidden}&=&(c_2(V_2)-\frac{1}{2}c_2({\cal M}))\int J \wedge J \wedge J
=4
\end{eqnarray}

Given eqs(\ref{mfunc}) one can determine 
\cite{CHOI,LUKAS} the soft supersymmetry breaking terms
for the 
observable sector 
gaugino masses $M_{1/2}$, scalar masses $m_0$ and trilinear scalar 
masses $A$
as functions of the auxiliary fields $F_S$ and $F_T$ of the moduli 
$S,T$ fields respectively.
\footnote{We assume very small 
$CP$-violating phases in the soft terms. This assumption 
is supported by the CP-structure 
of the soft terms in  modular invariant 
string theories when the moduli are at special algebraic 
points of the string moduli space\cite{us}.}
\begin{eqnarray}
M_{1/2}&=&
\frac{\sqrt{3}Cm_{3/2}}{(S+\bar{S}+\alpha(T+\bar{T})}
\Biggl((S+\bar{S})\sin\theta+\frac{\alpha(T+\bar{T})\cos\theta}
{\sqrt{3}}\Biggr), \nonumber \\
m_0^2&=&V_0+m_{3/2}^2-\frac{3 m_{3/2}^2 C^2}{3(S+\bar{S})+
\alpha(T+\bar{T})} \nonumber \\
&\times&\Biggl\{\alpha(T+\bar{T})\Biggl(2-
\frac{\alpha(T+\bar{T})}{3(S+\bar{S})+\alpha(T+
\bar{T})}\Biggr)\sin^2 \theta \nonumber \\
&+&(S+\bar{S})\Biggl(2-
\frac{3(S+\bar{S})}{3(S+\bar{S})+\alpha(T+
\bar{T})}\Biggr)\cos^2 \theta \nonumber \\
&-&\frac{2\sqrt{3}\alpha(T+\bar{T})(S+\bar{S})}
{3(S+\bar{S})+\alpha(T+\bar{T})}\sin\theta\cos\theta\Biggr\}, \nonumber \\
A&=&\sqrt{3}Cm_{3/2}\Biggl\{\Biggl(-1+
\frac{3\alpha(T+\bar{T})}{3(S+\bar{S})+\alpha(T+\bar{T})}\Biggr)
\sin\theta \nonumber \\
& &+\sqrt{3}\Biggl(-1+\frac{
3(S+\bar{S})}{3(S+\bar{S})+\alpha(T+
\bar{T})}\Biggr)\cos\theta\Biggr\}
\end{eqnarray}
while the  $B$-soft term associated with non-perturbatively generated 
$\mu$ term in the superpotential 
is given by \cite{BKL}:

\begin{eqnarray}
B_{\mu}&=&m_{3/2}\Bigl[-3 C \cos\theta -\sqrt{3}C\sin\theta  \nonumber \\
&+& \frac{6 C \cos\theta (S+\bar{S})}{3(S+\bar{S})+
\alpha (T+\bar{T})}  +\frac{2 \sqrt{3} C \sin\theta \alpha (T+\bar{T})}{
3(S+\bar{S})+\alpha (T+\bar{T})}-1 \Bigr] +
F^S \frac{\partial {ln \mu}}{\partial S}+F^T \frac{\partial {ln \mu}}
{\partial T} \nonumber \\
\label{beta}
\end{eqnarray}
In (2),(3) the auxiliary fields are parametrized as follows \cite{IBA:Spain}:
\begin{eqnarray}
F^S&=&\sqrt{3} m_{3/2} C (S+\bar{S}) \sin\theta, \nonumber \\
F^T&=&m_{3/2} C (T+\bar{T})\cos\theta
\end{eqnarray}
and $\theta$ is the goldstino angle which specifies the extent to which 
the supersymmetry breaking resides in the dilaton versus the moduli 
sector. Also $m_{3/2}$ is the gravitino mass 
and $C^2=1+\frac{V_0}{3 m_{3/2}^2}$ with $V_0$ the tree level vacuum 
energy density.
Note that in the limit $\alpha (T+\bar{T})\rightarrow 0$ we recover the soft
terms of the weakly coupled large $T$-limit of Calabi-Yau compactifications
\cite{IBA:Spain}.

\section{Supersymmetric particle spectrum and relic abundances of 
the neutralino}
We now consider the supersymmetric particle spectrum and the calculation 
of the relic abundance of the lightest neutralino. We also report the 
event rates for direct detection. We discuss the details of this calculation 
in section 4.

Our parameters are the goldstino angle $\theta$,
$\alpha(T+\bar{T})$,$sign \mu$ (which is not determined by the radiative 
electroweak symmetry breaking constraint), where 
$\mu$ is the Higgs mixing parameter in the 
low energy superpotential, 
and $\tan\beta $ (i.e the ratio of the two Higgs vacuum 
expectation values $\tan\beta=\frac{<H_2^0>}{<H_1^0>}$) if we leave  
$B$ a free parameter determined 
by  the minimization of the Higgs potential. 
If $B$ instead is given by
(\ref{beta}), one determines the value of $\tan\beta$. 
For this purpose  we take $\mu$ independent of $T$ and $S$
because of our lack of knowledge of 
$\mu$ in $M$-theory.
We also set $C=1$ 
in the above expressions assuming zero cosmological constant.
For the goldstino angle we choose three representative values:
$\theta=\frac{7\pi}{20},\frac{\pi}{4},\frac{\pi}{8}$.
The soft masses start running from 
a mass $R_{11}^{-1}\sim 7.5 \times 10^{15} GeV$ with 
$R_{11}$ the extra $M$-theory dimension. This is perhaps the most 
natural choice. However values as low as $10^{13}$ GeV are possible 
and have been advocated by some authors \cite{NANO}. 
However, the recent analysis of \cite{NILLES} disfavours such 
scenarios. For the most part of our analysis we shall consider the 
former value of $R_{11}$, but we shall also comment
on the consequences of the latter.

Then using (2) (3) as boundary conditions for the soft terms,
one evolves the renormalization 
group equations down to the weak scale and determines 
 the sparticle 
spectrum compatible with the constraints of correct electroweak symmetry 
breaking and experimental 
constraints on the sparticle 
spectrum from unsuccessful 
searches at LEP,Tevatron etc, 
and also that the LSP providing  a good dark matter candidate.

Electroweak symmetry breaking is characterized by the extrema equations 
\begin{eqnarray}
\frac{1}{2}M_Z^2&=&\frac{\bar{m}^2_{H_1}-\bar{m}^2_{H_2}\tan^2 \beta}
{\tan^2 \beta -1}-\mu^2 \nonumber \\
-B\mu&=&\frac{1}{2}(\bar{m}^2_{H_1}+\bar{m}^{2}_{H_2}+2\mu^2)\sin 2\beta
\end{eqnarray}
where 
\begin{equation}
\bar{m}^2_{H_1,H_2}\equiv m^2_{H_1,H_2}+\frac{\partial \Delta V}{
\partial {v^2_{1,2}}}
\end{equation}
and $\Delta V=(64 \pi^2)^{-1} {\rm {STr}} M^4[ln (M^2/Q^2)-\frac{3}{2}]$ 
is the 
one loop contribution to the Higgs effective potential. We include
contributions from the third generation of particles and sparticles.

Since $\mu^2 \gg M_Z^2$ for most of the allowed
region of the parameter space \cite{NATH}, the following 
approximate relationships 
hold at the 
electroweak scale for the 
masses of neutralinos and charginos, which of 
course depend on the details of 
electroweak symmetry breaking. 
\begin{eqnarray}
m_{\chi_1^{\pm}}\sim m_{\chi_2^0}\sim 2 m_{\chi_1^0} \nonumber \\
m_{\chi_{3,4}^{0}}\sim m_{\chi_2^{\pm}}\sim |\mu|
\label{neutralinos}
\end{eqnarray}
In (\ref{neutralinos}) $m_{\chi_{1,2}^{\pm}}$ are the chargino mass 
eigenstates and $m_{\chi_{i}^{0}},i=1\ldots 4$ are the four neutralino mass 
eigenstates with $i=1$ 
denoting the lightest. The former arise after diagonalization of the 
mass matrix. 
\begin{equation}
M_{ch}=\left(\begin{array}{cc}\\
M_2 & \sqrt{2} m_W \sin\beta \\
m_W \cos\beta & -\mu 
\end{array}\right)
\end{equation}

We focus on  the extreme $M$-theory limit 
\footnote{ Note that in this limit the dilaton dominated scenario ,
i.e ($\theta=\frac{\pi}{2}$) is not feasible.} in which 
$\alpha(T+\bar{T})=2$.
An interesting case is when 
the scalar 
masses are much smaller than the gaugino masses at the unification scale.
\footnote{Note that the weakly coupled case scalar masses 
are comparable to gaugino mass $m_{0} \sim M$ at the unification scale.}
For instance this can be achieved by a 
goldstino angle $\theta=\frac{7\pi}{20}$.
In this case sleptons are much lighter 
compared to the gluino than in the weakly coupled 
Calabi Yau compactifications of the heterotic
string for any value 
of $\theta$. As a result for high values of $\tan\beta$ \footnote{In case 
we determine  $B$ from the electroweak 
symmetry breaking $\tan\beta$ is a free parameter;
also in this case  
$M_{\tilde{g}}:m_{\tilde{L}_L}:m_{\tilde{e}_R}\sim
1:0.25:0.14$} there is a possibility 
that right handed selectrons or the lightest stau 
mass eigenstate become the 
LSP, where  
the stau mass matrix is given by the expression 

\begin{equation}
{\cal M}_{\tau}^2=\left(\begin{array}{cc} \\
{\cal M}_{11}^2 & m_{\tilde{\tau}}(A_{\tau}+\mu \tan\beta) \\
m_{\tilde{\tau}}(A_{\tau}+\mu \tan\beta) & {\cal M}_{22}^2
\end{array}\right)
\end{equation}
This of course is phenomenologically unacceptable
\cite{EHNOS,MATHEMA} and results in 
strong constraints in the sparticle spectrum. In fig.3 we plot the 
critical value of gravitino mass $m_{3/2}^{c}$ above, (below)  which 
$m_{\tilde{\tau}_2}$ or $m_{\tilde{e}_R}<m_{\chi_1^0}$ 
$(m_{\tilde{\nu}}<43GeV)$ versus $\tan\beta$. The acceptable 
parameter space lies between the upper and lower bound 
in fig.3. 
One can see that 
because of the above contraint 
$\tan\beta \leq 13$ in this M-theory limit because the acceptable 
parameter space vanishes for larger values of $\tan\beta$.
We also plot in fig.4 the critical chargino mass $m_{\chi_1^{\pm}}$ 
,above which the right handed seleptons become the LSP, versus 
$\tan\beta$. In the same figure we also draw the experimental lower 
bound
for the chargino mass from LEP of about $83$ GeV (horizontal line). 
The $\tan\beta$ dependence of these constraints may be understood from the 
$D$-term contribution to the $\tilde{e}_R$ and $\tilde{\nu}$ mass 
formulas
\begin{equation}
\tilde{m}^2_i=c_i m_{3/2}^{2}-d_i \frac{\tan^2 \beta-1}{\tan^2 \beta+1} 
M^2_W
\end{equation}
where the $c_i$ are some RGE-dependent constants and 
$d_{\tilde{e}_R}=-\tan^2 \theta_W<0$ whereas $d_{\tilde{\nu}}=
\frac{1}{2}(1+\tan^2 \theta_W)>0$.

In the allowed region of fig.1 the LSP is the lightest 
neutralino $\chi_{1}^{0}$.
Assuming $R$-parity conservation 
the LSP is stable  and consequently 
can provide a good dark matter candidate. 
It is a linear combination of the superpartners of the 
photon, $Z^0$ and neutral-Higgs bosons,
\begin{equation}
\chi_1^0=N_{11} \tilde{B}+N_{12}\tilde{W}^3+N_{13}\tilde{H}_1^0+N_{14}\tilde{H}_2^0
\end{equation}
The neutralino $4\times 4$ mass matrix can be written as
$$\left(\begin{array}{cccc} \\
M_1 & 0 & -M_Z A_{11} & M_Z A_{21} \\
0  & M_2 & M_Z A_{12} &-M_Z A_{22} \\
-M_Z A_{11} & M_Z A_{12} & 0 & \mu \\
M_Z A_{21} & -M_Z A_{22} & \mu & 0
\end{array}\right)$$ 
with 
$$\left(\begin{array}{cc} \\
 A_{11} & A_{12} \\
A_{21} & A_{22}\end{array}\right)= 
\left(\begin{array}{cc} \\
\sin\theta_{W} \cos\beta & \cos \theta_W \cos\beta \\
\sin\theta_W \sin\beta & \cos\theta_W \sin\beta 
\end{array}\right)$$

In fact the lightest 
neutralino in this model is almost a pure  bino ($\tilde {B}$),
which means $f_g\equiv |N_{11}|^2+|N_{21}|^2\gg 0.5$.
Most cosmological 
models predict that the relic abundance of neutralinos \cite{ARNO} 
satisfies
\begin{equation}
0.1\leq \Omega_{LSP} h^2 \leq 0.4
\label{COSMO}
\end{equation}

We calculated the relic abundance of the lightest neutralino using standard 
technology \cite{MARK} and found strong constraints on 
the resulting spectrum. In particular  for $\mu<0$ the lower limit on the 
relic abundance  results in 
a lower limit on the gravitino mass of about 200 GeV. In figs(5-7). 
we plot the 
relic abundance of the lightest neutralino versus the gravitino 
mass for various values of 
$\tan\beta$ for goldstino angle of $\theta=\frac{7\pi}{20}$. We also note 
that for the allowed parameter space $\Omega_{LSP} h^2$ never 
exceeds the upper limit of 0.4. However, for other values of the goldstino 
angle the upper limit on $\Omega_{LSP} h^2$ 
can constrain the gravitino mass. 
The lower limit also constrains the allowed values of the $\tan\beta$ 
parameter even further. For instance one can see from the plots 
that $\tan\beta<10$ in order that $\Omega_{LSP} h^2 \geq 0.1$.
In the allowed physical region direct 
detection rates are of order $10^{-2}-10^{-4} events/Kg/day$. 
The lightest Higgs mass $m_h$ is in the range 
$87GeV\leq m_h \leq 115 GeV$, while the neutralino mass is in the 
range $77GeV \leq m_{\chi_0} \leq 195 GeV$. 
 At this point 
is worth    
 mentioning
that if one chooses to run 
the soft masses from the mass 
$R_{11}^{-1}\sim O(10^{13})GeV$ instead of 
$7.5 \times 10^{15} GeV$ the cosmological constraint (
\ref{COSMO}) is very powerful and 
eliminates all of the parameter space since the relic abundance is always 
much smaller than 0.1, when $\theta=\frac{7\pi}{20}$.
For $\mu>0$ the maximum gravitino mass 
above which $m_{{\tilde e}_R}<m_{\chi_1^0}$
 or $m_{{\tilde \tau}_2}<m_{\chi_1^0}$ is smaller for fixed $\tan\beta$.
Clearly this novel $M$-theory limit provide us with a phenomenology 
distinct from the weakly coupled case \footnote{Note however, that this 
M-theory limit is somewhat similar to  the O-I orbifold model 
\cite{IBA:Spain,BK} and for a particular 
goldstino angle , different from the dilaton-dominated limit. 
On the other hand the O-I model has non-universal soft supersymmetry-
breaking terms at the string scale.}  
which should be a subject of experimental scrutiny.

For other values of the goldstino angle 
for which scalar masses are comparable to the 
gaugino masses (a case which is more similar to 
the weakly coupled limit \cite{IBA:Spain} in which 
$m_0\sim \frac{1}{\sqrt{3}} M_{1/2}$) we do not obtain constraints from
the bounds on the mass of  
right handed selectrons  and staus, but in this case the upper 
limit on the relic 
abundance 
leads to an upper limit on  the gravitino mass. For instance, for a
goldstino angle of $\theta=\frac{\pi}{4}$ (see fig.7) and 
$\tan\beta=2.5$, the requirement that 
$\Omega_{LSP} h^2 \leq 0.4$ results in $m_{3/2}\leq 365$ GeV. This results in 
an upper limit for the lightest Higgs mass 
$m_h\leq 100 GeV$. The lower limit 
is now $m_{3/2}\geq 115$ GeV. However, the LEP limit on the 
chargino mass of 82 GeV requires that $m_{3/2}\geq 150$ GeV. In this 
case the lightest neutralino is in the range 
$52 GeV\leq m_{\chi_1^0} \leq 148 GeV for  \mu<0$. For $\mu>0$ we have 
$65 GeV \leq m_{\chi_1^0} \leq 153 GeV$. Detection rates of the LSP 
for $^{73} Ge$ detector are in the range 
$6 \times 10^{-2}-10^{-4}events/Kg/day$ for $\mu<0$ and $O(10^{-3})-O(10^{-5}$
events/Kg/day for $\mu>0$. For higher $\tan\beta$ values one can obtain 
higher detection rates.

For a goldstino angle $\theta=\frac{\pi}{8}, \tan\beta=2.5$, the upper limit 
on the relic abundance leads to an upper limit on the gravitino mass
$m_{3/2}\leq 360$ GeV. As a result $m_{\chi_0^1}\leq 119$GeV and 
$m_{\chi_1^{\pm}}\leq 220$ GeV for $\mu<0$. Similarly, $m_h\leq  105$GeV.
For higher values of $\tan\beta$, i.e $\tan\beta=15$ the neutralino mass 
is $60\leq m_{\chi_1^0} \leq 120$ GeV and $m_{h}\leq 125$ GeV. The detection
rates have been calculated in three different nuclei $^{73}Ge,^{131}Xe,
^{208}Pb$ and are in the range of order $10^{-1}$ to 
order $10^{-4} events/Kg/day$. The higher the atomic number of the target 
nuclei the higher the detection rate since the strength of the
scalar interaction is sensitive to the atomic number (see Fig.12 and 
the discussion in section 4). The upper 
line corresponds to $^{208}Pb$ and the lower to $^{73}Ge$. Also the 
higher detection rates occur for low LSP mass ($
m_{\chi_1^0}\sim 60 GeV$) since for higher LSP mass the 
LSP becomes more and more Bino and the interaction is suppressed.
The reader should bear in mind the important relationship $m_{\chi_1^0}\sim 
\frac{1}{2}m_{\chi_{1}^{\pm}}$ (\ref{neutralinos}). 
That relationship (\ref{neutralinos}) 
can provide an important consistency check if a positive 
signal with $m_{\chi_1^0}\leq 40$ GeV is found.
We also calculated the modulation effect 
in the total event rate which is small ($\leq 0.05$).

In summary, we have analyzed the supersymmetric 
spectrum and the properties 
of the lightest neutralino (LSP) in the 11-dimensional limit of M-theory
for the extreme M-theory case in which $\alpha(T+\bar{T})=2$. 
The most striking result 
,in the case of small scalar masses compare with  gluino masses, is that one 
obtains a limit on $\tan\beta\leq 13$, since above that value the 
right handed selectron or the lightest stau is the LSP, which is 
phenomenologically unacceptable 
since the LSP should be electrically neutral \cite{EHNOS,MATHEMA}.
Also the cosmological constraint on the relic abundance of the LSP results
in a lower limit on the gravitino mass $m_{3/2}\geq 200$GeV. 
This further constrains  $\tan\beta$; $\tan\beta<10$.
In this case 
the upper limit on the relic abundance is not relevant since 
$\Omega_{LSP} h^2 < 0.4$ for goldstino angle $\theta=\frac{7\pi}{20}$ 
and for all the allowed values of $\tan\beta$. Also the lower 
limit on the relic abundance excludes the case of 
$R_{11}^{-1}\sim O(10^{13})$ GeV.
The scenario with $B_{\mu}$ given by (\ref{beta}),
and $\mu$ independent of, $S$  and, $T$ is excluded since one has 
to go to non-perturbative Yukawa couplings in order to obtain a value of 
$\tan\beta$ consistent with $B_{\mu}$ as in  (\ref{beta}).
For a goldstino angle of $\theta=\frac{(7.2)\pi}{20}$ the maximum value 
of gravitino mass for a fixed $\tan\beta$ is much lower (for $\mu>0$ the 
parameter space vanishes) from the requirement that the 
lightest neutralino is 
the LSP and $\tan\beta(max) \sim 12$. 
For other values of the goldstino angle 
(which resemble more the weakly coupled 
Calabi-Yau compactifications for which 
$m_{0}=\frac{1}{\sqrt{3}}M_{1/2}$) the upper bound on the relic 
abundance results in an upper bound on the gravitino mass. For 
a goldstino angle, $\theta=\frac{\pi}{4}$ 
and $\tan\beta=2.5$ then we find $m_{3/2}\leq 365$GeV. 
For $\tan\beta=12.5$ the upper limit on the relic abundance 
leads to an upper limit on the LSP mass of $\sim 160$ GeV.
Direct detection rates of the lightest neutralino are in 
the range of $10^{-1}-10^{-4} events/Kg/day$.

\section{Direct detection}

Detection of the LSP is  a field in which particle physics, nuclear physics 
and cosmology play important role in the analysis. Although there are 
uncertainties which stem from all the above mentioned disciplines the 
main uncertainty arises from ignorance of the supersymmetric 
parameters involved. 
A neutral LSP cannot be detected in collider experiments because of its 
stability. Experimental detection is therefore focused on underground 
detectors. There are two main methods,
direct  detection of the nuclear recoil of order $(10 KeV)$ induced 
by the weak interaction of the LSP with the nucleus
\cite{WIGM} and indirect detection which measures
the upward muon flux from energetic neutrinos of order 100 GeV 
resulted from capture of LSP in massive objects like the sun or earth
\cite{INDI}.

The typical cross section involved 
is $\sigma_0\sim 10^{-38} cm^2$, and for an LSP in the 
halo with a typical mass of 100GeV 
interacting with a $^{76} Ge$ nucleus of mass 67.9 GeV and moving with 
a velocity $v$ of 300Km/sec this leads to a 
detection rate of $\sim 1 event/Kg/year$. 
In what follows we will describe in detail the microscopic interactions of 
LSP with matter and the process of translating these interactions to the 
nuclear level\cite{GR,ELLIS,BOTTINO,DREES,
RUS,AN,GORDON,SWE,VER,BK,KANE,BKL,JOHN,KHALIL}. Because 
the neutralino is a Majorana particle 
it does not lead to vector interactions with matter because of the 
identity
\begin{equation}
\bar{\chi_1^0} \gamma^{\mu} \chi_1^0=0
\end{equation}
This property is important since it makes the observation of LSP 
a very difficult experimental task.
The 
fundamental interaction of $\chi_1^0$ with quarks involves three possibilities:
exchange of $Z$ boson, s-quark exchange and Higgs exchange.
The total interaction is of the form
\begin{equation}
{\cal L}_{eff}={\cal L}_{eff}^{(SD)}+{\cal L}_{eff}^{(SI)}
\end{equation}
where ${\cal L}_{eff}^{(SD)}$ is a spin dependent interaction and 
${\cal L}_{eff}^{(SI)}$ is a spin independent part. 
In the following we describe in detail the 
above mentioned three possibilities.

\subsection{The $Z$-exchange contribution}
This can arise from the interaction of Higgsinos with $Z$. The relevant 
microscopic Lagrangian reads:
\begin{equation}
{\cal L}=\frac{g}{\cos\theta_W}\frac{1}{4}(|N_{31}|^2-|N_{41}|^2)
\bar{\chi_1^0} \gamma_{\mu}\gamma_5 \chi_1^0 Z^{\mu}
\label{zex}
\end{equation}
\begin{figure}
\epsfxsize=3.25in
\epsfysize=3.65in
\epsffile{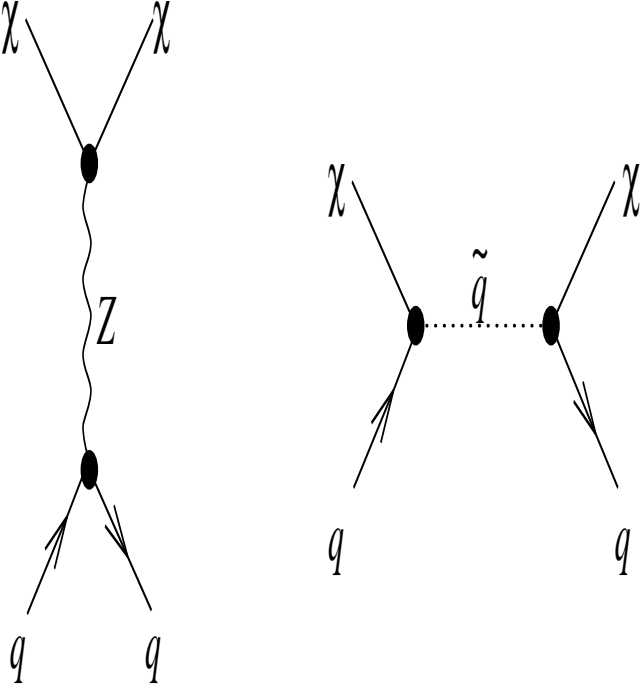}
\caption{Spin dependent scattering of $\chi$ from quarks}
\end{figure}
which leads to the effective 4-fermion interaction
\begin{equation}
{\cal L}_{eff}=\frac{g}{\cos\theta_W}\frac{1}{4} 2 (
|N_{31}|^2-|N_{41}|^2)(-\frac{g}{2 \cos\theta_W}
\frac{1}{q^2- M_Z^2} \bar{\chi_1^0} \gamma^{\mu}\gamma_5 \chi_1^0)
J_{\mu}^{Z}
\end{equation}
where $N_{31}$ and $N_{41}$ are defined as in Eq.(15).
The neutral hadronic current $J_{\lambda}^Z$ is given by
\begin{equation}
J_{\lambda}^Z=-\bar{q}\gamma_{\lambda}\Bigl\{
\frac{1}{3}\sin^2 \theta_W-\Bigl[\frac{1}{2}(1-\gamma_5)-
\sin^2 \theta_W\Bigr]\tau_3\Bigr\}q
\end{equation}
at the nucleon level it can be written as
\begin{equation}
\tilde{J}_{\lambda}^Z=-\bar{N}\gamma_{\lambda}\Bigl\{
\sin^2 \theta_W-g_V(\frac{1}{2}-\sin^2 \theta_W)\tau_3+
\frac{1}{2}g_A \gamma_5 \tau_3\Bigr\}N
\end{equation}
Consequently the effective Lagrangian can be written
\begin{equation}
{\cal L}_{eff}=-\frac{G_F}{\sqrt{2}}(\bar{\chi_1^0} \gamma^{
\nu}\gamma^5 \chi_1^0)J_{\nu}(Z)
\end{equation}
where 
\begin{equation}
J_{\nu}(Z)=\bar{N}\gamma_{\lambda}[
f_V^0(Z)+f_V^1(Z) \tau_3+f_A^0  (Z) \gamma_5+
f_A^1(Z) \gamma_5 \tau_3]N
\end{equation}
and 
\begin{eqnarray}
f_V^0(Z)&=&2(|N_{31}|^2-|N_{41}|^2)\frac{M^2_Z}{M^2_Z-q^2}\sin^2 \theta_W
\nonumber \\
f_V^1(Z)&=&-2(|N_{31}|^2-|N_{41}|^2)\frac{M^2_Z}{M^2_Z-q^2}g_V(
\frac{1}{2}-\sin^2\theta_W) \nonumber \\
f_A^0(Z)&=&0 \nonumber \\
f_A^1(Z)&=&2(|N_{31}|^2-|N_{41}|^2)\frac{M^2_Z}{M^2_Z-q^2}\frac{1}{2}g_A
\end{eqnarray}
with $g_V=1.0,g_A=1.24$.  
For the Z-exchange (\ref{zex}) we see that this exchange not only 
depends on the Higgsino component of the 
neutralino  but on their mismatch as well.
This interaction is suppressed for an LSP 
which is almost Bino i.e 
$\chi_1^0 \sim \tilde {B}$, 
since  the mixing couplings $N_{31},N_{41}$ 
are small. Nevertheless  in some regions 
of the parameter space $N_{31}\sim 0.2$  and  
$N_{41}\sim 0.01$  and therefore this interaction is not 
entirely negligible although much smaller than the scalar interaction.

\subsection{The squark mediated interaction}
The effective Lagrangian at the nucleon level in this case is 
given \cite{DREES} by
\begin{equation}
{\cal L}_{eff}=-\frac{G_F}{\sqrt{2}}(
\bar{\chi}^0_{1}\gamma^{\lambda}\gamma^5 \chi^0_1)J_{\lambda}(\tilde{q})
\end{equation}
\begin{equation}
J_{\lambda}(\tilde{q})=\bar{N}\gamma_{\lambda}
{f_V^0(\tilde{q})+f_V^1(\tilde{q})\tau_3+f_A^0(\tilde{q})
\gamma_5+f_A^1(\tilde{q})\gamma_5 \tau_3}N
\end{equation}
with 
\begin{eqnarray}
f_V^0 &=& 6(\beta_{0R}-\beta_{0L}),\;\;
f_V^1=2(\beta_{3R}-\beta_{3L}) \nonumber \\
f_A^0 &=& 2g_V(\beta_{0R}+\beta_{0L}),\;\;f_A^1=2 g_A (\beta_{3R}+\beta_{3L})
\end{eqnarray}
with
\begin{eqnarray}
\beta_{0R} &=&(\frac{4}{9}\chi^2_{\tilde{u}_R}+
\frac{1}{9}\chi^2_{\tilde{d}_R}|N_{11}\tan\theta_W|^2 \nonumber \\
\beta_{3R}&=&(\frac{4}{9}\chi^2_{\tilde{u}_R}-
\frac{1}{9}\chi^2_{\tilde{d}_R}|N_{11}\tan\theta_W|^2 \nonumber \\
\beta_{0L}&=&|\frac{1}{6}N_{11}\tan\theta_W+\frac{1}{2}
N_{21}|^2\chi^2_{\tilde{u}_L}+
|\frac{1}{6}N_{11}\tan\theta_W-
\frac{1}{2}
N_{21}|^2\chi^2_{\tilde{d}_L} \nonumber \\
\beta_{3L}&=&|\frac{1}{6}N_{11}\tan\theta_W+\frac{1}{2}
N_{21}|^2\chi^2_{\tilde{u}_L}-
|\frac{1}{6}N_{11}\tan\theta_W-
\frac{1}{2}
N_{21}|^2\chi^2_{\tilde{d}_L} 
\end{eqnarray}
with
\begin{eqnarray}
\chi^2_{qL}&=&\cos\theta_{\tilde{q}}^2 \frac{M_W^2}{m_{
\tilde{q_1}^2}-q^2}+\sin\theta^2_{\tilde{q}}\frac{
M_W^2}{m_{\tilde{q_2}^2}-q^2} \nonumber \\
\chi^2_{qR}&=&\sin\theta^2_{\tilde{q}}\frac{
M_W^2}{m_{\tilde{q_1}^2}-q^2}+
\cos\theta_{\tilde{q}}^2 \frac{M_W^2}{m_{
\tilde{q_2}^2}-q^2}
\end{eqnarray}
where $\tilde{q}_{1,2}$ are the squark mass eigenstates 
and $\theta_{\tilde{q}}$  the squark mass matrices mixing angles.

\subsection{The intermediate Higgs contribution}
The scalar interaction receives contributions from 
Higgs exchange and from the squark exchange. We first describe 
the Higgs exchange. 
\begin{eqnarray}
{\cal L}_{H\chi_1^0\chi_1^0}&=&\frac{g}{\sqrt{2}}
((N_{21}-\tan\theta_W N_{11})N_{41}
\bar{\chi}^0_{1R}\chi_{1L}^0 H_2^{0*} \nonumber \\
&-&(N_{21}-
\tan\theta_W N_{11})N_{31} \bar{\chi}^0_{1R}\chi_{1L}^0 H_1^{0*})
+{\rm h.c}
\end{eqnarray}
\begin{figure}
\epsfxsize=3.25in
\epsfysize=3.65in
\epsffile{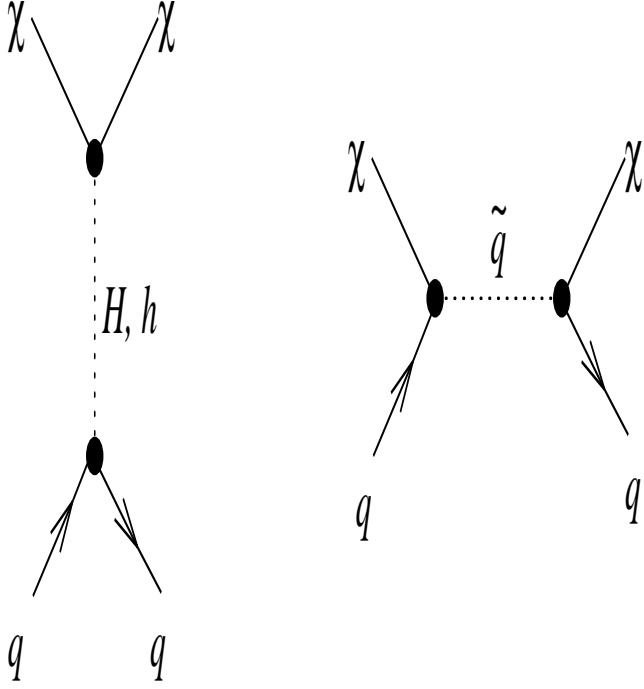}
\caption{Diagrams contributing to the scalar scattering of $\chi$ with
quarks}
\end{figure}
If we express the fields $H_1^{0*},H_2^{0*}$ in terms of the 
mass eigenstates $h,H,A$ we get \footnote{the term which 
contains $A$ will be neglected since it yields only a pseudoscalar 
coupling which does not lead to coherence}
\begin{equation}
{\cal L}_{eff}=-\frac{G_F}{\sqrt{2}} 
\bar{\chi}_1^0 \chi_1^0 \bar{N}[f_s^0(H)+f_s^1(H)\tau_3]N
\end{equation}
where 
\begin{equation}
f_s^0(H)=\frac{1}{2}(g_u+g_d)+g_s+g_c+g_b+g_t
\end{equation}
\begin{equation}
f_s^1(H)=\frac{1}{2}(g_u-g_d)
\end{equation}
with 
\begin{equation}
g_{a_i}=\Bigl[g_1(h)\frac{\cos \alpha}{\sin\beta}+
g_2(H)\frac{\sin \alpha}{\sin\beta}\Bigr]\frac{m_{a_i}}{m_N},
a_i=u,c,t
\end{equation}
\begin{equation}
g_{f_i}=\Bigl[-g_1(h)\frac{\sin \alpha}{\cos\beta}+
g_2(H)\frac{\cos \alpha}{\cos\beta}\Bigr]\frac{m_{f_i
}}{m_N},\;f_i=d,s,b
\end{equation}
\begin{equation}
g_1(h)=4(N_{11}\tan\theta_W-N_{21})(N_{41}\cos\alpha+
N_{31}\sin\alpha)\frac{m_N M_W}{m^2_h-q^2}
\end{equation}
\begin{equation}
g_2(H)=4(N_{11}\tan\theta_W-N_{21})(N_{41}\sin\alpha-
N_{31}\cos\alpha)\frac{m_N M_W}{m_H^2-q^2}
\end{equation}
where $m_N$ is the nucleon mass, $m_h,m_H$ are the one-loop 
corrected lightest and heavier Higgs CP-even mass eigenstates and
$\alpha$ is the Higgs mixing angle which at tree level is given by
\begin{equation}
\cos {2\alpha}=-\cos 2\beta\Biggl(\frac{m^2_A-M^2_Z}{m^2_H-m^2_h}
\Biggr)
\label{MIAN}
\end{equation}
However, we determine the Higgs mixing angle numerically by 
diagonalizing the one-loop $CP$-even Higgs mass matrix since as 
has been emphasized in \cite{AN} the determination of the 
Higgs mixing angle by inserting the one-loop values in the tree-level 
expression (\ref{MIAN}) leads to erroneous results.

We note that the scalar interaction involves an interference 
of the gaugino and the Higgsino components. Thus if the LSP were 
to be strictly a Bino or a Higgsino
, the scalar interaction would vanish and the 
neutralino-nucleus scattering would be strictly governed by the 
spin dependent interaction. However, such a situation is never 
realized and the scalar interaction plays the dominant role in the 
event rate analysis of the effective supergravity from $M$-theory.
At this point we must emphasize that the coherent interaction 
depends very much on the nuclear model. 
If only the up and down quarks contribute to the nucleon mass 
and  $m_u=5 MeV,m_d=10 MeV$  one finds \cite{addler}
\begin{equation}
f_s^0=1.86 (g_u+g_d)/2, f_s^1=0.49 (g_u-g_d)/2
\end{equation}
In this case the interaction is small.
However, if heavier quarks are also involved due to QCD effects, one 
obtains the pseudoscalar Higgs-nucleon interaction $f_s^0(H)$ by 
using effective quark masses as follows
\begin{equation}
m_u\rightarrow f_u m_N,\;\;m_d\rightarrow f_d m_N,m_s\rightarrow f_s m_N
,m_{Q}\rightarrow f_Q m_N\;\; {\rm for} \;{\rm heavy}\;{\rm
quarks}\;\; (c,b,t)
\end{equation}
where $m_N$ is the nucleon mass. The isovector contribution is 
now negligible. The parameters $f$ can be obtained by chiral symmetry 
breaking terms in relation to phase shift and dispersion analysis.
Following Cheng we obtain
\begin{equation}
f_u=0.021,\;f_d=0.037,\;f_s=0.140 
\end{equation}
As pointed by Shifman, Vainstein, and Zakharov,  
heavy quarks contribute to the mass of 
the nucleon through the anomaly \cite{SVK}.
Under the heavy quark expansion, the following substitution can be made 
for the heavy quarks $Q=c,b,t$, in a nucleon matrix element \cite{SVK}
\begin{equation}
m_Q \bar{Q} Q\rightarrow \frac{-2 \alpha_S}{24\pi} GG
\end{equation}
so we find that for each of the heavy quarks, $Q=c,b,t$ \cite{MARK}, 
\begin{equation}
<N|m_Q \bar{Q} Q|N>=\frac{2}{27}m_N\Bigl[1-\sum_{q=u,
d,s}f^{N}_{q}\Bigr]
\end{equation}
with $G$ the QCD field strength.

In the case of the scalar interaction,
the sum over all quarks in the nucleus 
produces a 
mass factor $A$, which enhances the contribution of the scalar term. 
Thus the heavier the nucleus the more sensitive it is to the 
spin-independent scattering.

We can also a have a small 
scalar contribution from the squark exchange, as was
first observed by Griest \cite{GR}, in which the effective Lagrangian 
at the nucleon level is
\begin{equation}
{\cal L}=\frac{G_F}{\sqrt{2}}[f_S^0(\tilde{q})\bar{N}N+
f_s^1(\tilde{q})\bar{N}\tau_3 N]\bar{\chi}_1^0\chi_1^0
\end{equation}
with $f_s^0(\tilde{q})=1.86\beta_{+}$ and $f_s^1(\tilde{q})=0.48
\beta_{-}$ \cite{addler}
with 
\begin{eqnarray}
\beta_{\pm}&=&\frac{1}{3}\tan\theta_W N_{11}\{2 \sin 2\theta_{\tilde{u}}
[\frac{1}{6}N_{11}\tan\theta_W+\frac{1}{2}N_{21}]\Delta_{\tilde{u}} 
\nonumber \\
&\mp & \sin 2\theta_{\tilde d}[\frac{1}{6}
N_{11}\tan\theta_W-\frac{1}{2}N_{21}]\Delta_{\tilde{d}}\}
\end{eqnarray}

\subsection{The nuclear matrix elements}
Combining the results of the previous subsections we can write
\begin{equation}
{\cal L}_{eff}=-\frac{G_F}{\sqrt{2}}\Bigl\{
(\bar{\chi}_1^0\gamma^{\lambda}\gamma_5 \chi_1^0)J_{\lambda}+
(\bar{\chi}_1^0\chi_1^0)J\Bigr\}
\end{equation}
where 
\begin{equation}
J_{\lambda}=\bar{N}\gamma_{\lambda}(f_V^0+f_V^1 \tau_3+
f_A^0 \gamma_5+ f_A^1 \gamma_5 \tau_3)N
\end{equation}
with 
\begin{eqnarray}
f_V^0&=&f_V^0(Z)+f_V^0(\tilde{q}),\;\;f_V^1=f_V^1(Z)+f_V^1(\tilde{q})
\nonumber \\
f_A^0&=&f_A^0(Z)+f_A^0(\tilde{q}),\;\;f_A^1=f_A^1(Z)+f_A^1(\tilde{q})
\end{eqnarray}
and the scalar interaction is given by
\begin{equation}
J=\bar{N}(f_s^0+f_s^1 \tau_3)N
\end{equation}

The total cross section can be cast in the form \cite{MARK,JOHN}
\footnote{Note that form factors have been introduced in 
(\ref{Cross}) to take into account the fact that when the LSP 
and the detector nuclei have masses in the $100$GeV range 
the scattering becomes more complicated because, though the 
energy transferred to the nucleus is still very small, the 
three-momentum transfer $q\equiv |\bf{q}|\leq 2 \mu v$ ( 
where $\mu$ is the reduced mass and $v\sim 10^{-3} c$ is the 
average galactic neutralino velocity) can be larger than the 
inverse size of the nucleus \cite{ENGEL}.}

\begin{eqnarray}
\sigma &=&\sigma_0 (\frac{m_{\chi_1^0}}{m_N})^2 
(\frac{1}{(1+\eta)^2}\Bigl\{A^2 [[ \beta^2 (f_V^0-f_V^1 \frac{A-2Z}{A})^2
\nonumber \\
&+&(f_s^0-f_s^1 \frac{A-2 Z}{A})^2 ]I_0 (u_0)-
\frac{\beta^2}{2}\frac{2 \eta+1}{(1+\eta)^2}(f_V^0-f_V^1 
\frac{A-2 Z}{A})^2 I_1 (u_0)]
\nonumber \\
&+&(f_A^0 \Omega_0(0))^2 S_{00}(u_0)+
2f_A^0 f_A^1 \Omega_0(0)\Omega_1(0) S_{01}(u_0) \nonumber \\
&+&(f_A^1 \Omega_1(0))^2 S_{11}(u_0)\Bigr\}
\label{Cross}
\end{eqnarray}
where 
\begin{equation}
\sigma_0=\frac{1}{2\pi}(G_F m_N)^2\sim 0.77\times 10^{-38} cm^2
\end{equation}
and 
\begin{equation}
\eta=\frac{m_{\chi^0_1}}{m_N A};q_0=\beta\frac{2m_{\chi^0_1} c}{1+\eta}
\end{equation}
and the dimensionless quantity $u_0$ is defined as \cite{JOHN}
\begin{equation}
u_0=q_0^2 b^2/2
\end{equation}
with $b\sim 1.0 A^{1/6} \rm {fm}$ and $A=Z+N$ is the atomic number of 
the nuclei.
The form factors entering  (\ref{Cross}) are defined as 
\begin{equation}
I_{\rho}=(1+\rho)u_0^{-(1+\rho)}\int_{0}^{\infty} x^{1+\rho}
|F(x)|^2 dx,\;\;\rho=0,1
\end{equation}
where $F(q)$ is the nuclear form factor.
Another  appropriate form factor 
is the Woods-Saxon form factor \cite{ENGEL} 
 
\begin{equation}
I_{WS}=\frac{3j_1(qR_0)}{qR_0}exp[-\frac{1}{2}(qs)^2]
\end{equation}
or using appropriate expressions for the form factors 
\cite{JDV}(in a harmonic 
oscillator basis with size parameter $b$)  
\begin{equation}
I_{\rho}=\frac{1}{A^2}\Bigl\{Z^2 I_{ZZ}^{(\rho)}(u_0)+
2NZ I_{NZ}^{(\rho)}(u_0)+N^2 I_{NN}^{(\rho)}(u_0)\Bigr\}
\label{Form}
\end{equation}
where
\begin{eqnarray}
I_{\alpha\beta}^{(\rho)}(u_0)&=&
(1+\rho)\;\sum_{\lambda=0}^{N_{max(\alpha)}} 
\sum_{\nu=0}^{N_{max(\beta)}}\frac{\theta_{\lambda}^{(\alpha)}}{
\alpha}\frac{\theta_{\nu}^{(\beta)}}{\beta}
\frac{2^{\lambda+\nu+\rho}(\lambda+\nu+\rho)!}{u_0^{1+\rho}} \nonumber \\
&\times&[1-e^{-u_0}\sum_{k=0}^{\lambda+\nu+\rho} \frac{u_0^{\kappa}}
{\kappa!}]
\end{eqnarray}
We exhibit the dependence of the scalar form factors $I_0$ on the LSP mass 
in fig. (8)-(10) for three different nuclei.
 
Since the LSP is moving with velocity $v_Z$ with respect to the 
detector nuclei the detection rate for a target with mass $m$ is given 
by
\begin{equation}
R=\frac{d N}{dt}=\frac{\rho_0}{m_{\chi_1^0}}\frac{m}{A m_p}\int f(\bf{v})|v_z|
\sigma(|\bf{v}|)d^3 \bf v
\end{equation}
where $\rho_0=0.3 GeV/cm^3$ is the local LSP density in our vicinity which 
has to be consistent with a Maxwell distribution
\begin{equation}
f(v^{\prime})d^3 v^{\prime}=
(\sqrt{\pi} v_0)^{(-3)} e^{(-v^{\prime}/v_0)^2}d^3 v^{\prime}
\label{OKTO}
\end{equation}
and 
\begin{equation}
v_0=\sqrt{(2/3)<v^2>}=220Km/s
\end{equation}

More realistically, one should take into  account the motion of the 
Sun and Earth. This increases the total event rate and gives 
rise to a yearly modulation in the event rate which might serve 
as a method of distinguishing signal from noise if many events 
are found \cite{FREESE}. Following Freese, Frieman and Gould 
and Sadoulet \cite{FREESE}, one subtracts the Earth velocity 
$\bf{v_E}$ from $v^{\prime}$ in (\ref{OKTO}) to get the 
velocity $v$ of the LSP in the Earth frame.
\begin{equation}
{{\bf v^{\prime}}=\bf{v}+\bf{v_E}},\;\;{v^{\prime}}^2=v^2+
v^2_E+2 v v_E \cos\chi
\end{equation}
where $\chi$ is the angle between the LSP velocity in the Earth frame 
and the direction of the Earth's motion.
As a function of time, $v_E$ changes as the Earth's motion comes 
into and out of alignment and the event rate peaks 
around second of June. This is taken into account using
\begin{equation}
v_E=v_0\Bigl[1.05+0.07 \cos \Bigl(\frac{2\pi (t-t_p)}{\rm{1yr}}\Bigr)
\Bigr]
\end{equation}
where $t_p=\rm{June} 2^{nd}\pm 1.3 days$. Changing variables, and 
integrating over angles one obtains
\begin{equation}
f_1(v)dv=\frac{v dv}{v_E v_0 \sqrt{\pi}}
\Bigl\{e^{-\frac{(v-v_e)^2}{v_0^2}}-
e^{-\frac{(v+v_E)^2}{v_0^2}}\Bigr\}
\end{equation}
Assuming a Gaussian parametrization for the form factors 
\cite{ELLIS} and averaging over the Maxwellian distribution of 
dark matter velocities the total scattering rate is given by
\begin{eqnarray}
R&=&(R_{S.I}+R_{S.D}) \frac{4 m_{\chi_1^0} m_A}{(m_{\chi_1^0}+m_A)^2} \nonumber \\
&\times&\frac{\rho_0}{0.3 GeV cm^{-3}}
\frac{|V_E|}{320 km s^{-1}}\frac{events}{Kg d}
\end{eqnarray}
where the 
\begin{eqnarray}
R_{S.I}&=&840 m_A^2 M_Z^4 \zeta |A_{S.I}|^2
\end{eqnarray}
and 
\begin{eqnarray}
\zeta&=&\frac{0.573}{b}\Bigl(1-\frac{exp[-b/(1+b)]}{\sqrt{1+b}}
\frac{\rm{erf}(1/\sqrt{1+b})}{\rm{erf}(1)}\Bigr)
\end{eqnarray}
is a form factor with 
\begin{equation}
b=\frac{m_{\chi_1^0}^2 m_A^2}{((m_{\chi_1^0}+m_A)^2}\frac{8}{9}\sigma^2
r^2_{charge}
\end{equation}
where $\sigma=0.9\;10^{-3}$ is the velocity dispersion and 
$r_{charge}=(0.3+0.89A^{1/3})\rm{fm}$.

\section{Conclusions}
In this work we have analysed the supersymmetric particle spectrum 
of the 
extreme M-theory case 
$(\alpha(T+\bar{T})=2)$ with the requirement 
that the lightest supersymmetric particle 
is the dominant component of the dark matter in the universe.
We also calculated in detail the detection rates of direct interaction 
of neutralinos with various nuclei $^{131} Xe, ^{73}Ge, ^{208}Pb$. The 
LSP in the effective supergravity from 
M-theory is almost a Bino. This has the important 
effect that the scalar component (through Higgs exchange) is the dominant
interaction of the LSPs with terrestrial nuclei. The strength of this 
interaction is such that this should be detectable 
in the near future by nuclei with {\it large atomic number} $A$, since 
these nuclei are more sensitive to the scalar interaction as illustrated 
in Fig.12. We obtain maximum event rates of order $ 50 events/Kg/year$ for 
large values of $\tan\beta$ and lightest chargino mass of order $100$ GeV.
The detection rates depend also on the sign of the $\mu$ parameter as well 
the nucleon model . 
Therefore the 
allowed  parameter space can be limited by experimental investigation.
We also calculated the magnitude of the modulation effect due to the annual 
revolution of Earth which is small ($\leq 0.05$).  

In the  scenario with small scalar masses compare to 
gaugino masses, i.e $m_0 \ll M_{1/2}$ the 
parameter $\tan\beta$ is constrained to be small, i.e  $\tan\beta \leq 12$
from the requirement 
that the LSP should be the lightest neutralino. After imposing the 
limits on the relic abundance of the lightest neutralino the upper 
limit on $\tan\beta$ becomes smaller ($\tan\beta \leq 10$).

For scenarios with scalar masses comparable to the gaugino masses at the 
unification scale the upper limit from cosmology on the relic abundance 
of the LSP imposes an upper bound on the gravitino mass and gluino masses. 

The above results motivate the investigation of the effect of 
CP violating phases in the soft supersymmetry-breaking terms on the detection
rates as well as the investigation of the size of the modulation effect on the 
differential event rate. The above will be the
subject of a future publication \cite{GVK}.

\section*{Acknowledgements}
This research is supported in part by PPARC.

\newpage

\begin{figure}
\epsfxsize=6in
\epsfysize=8.3in
\epsffile{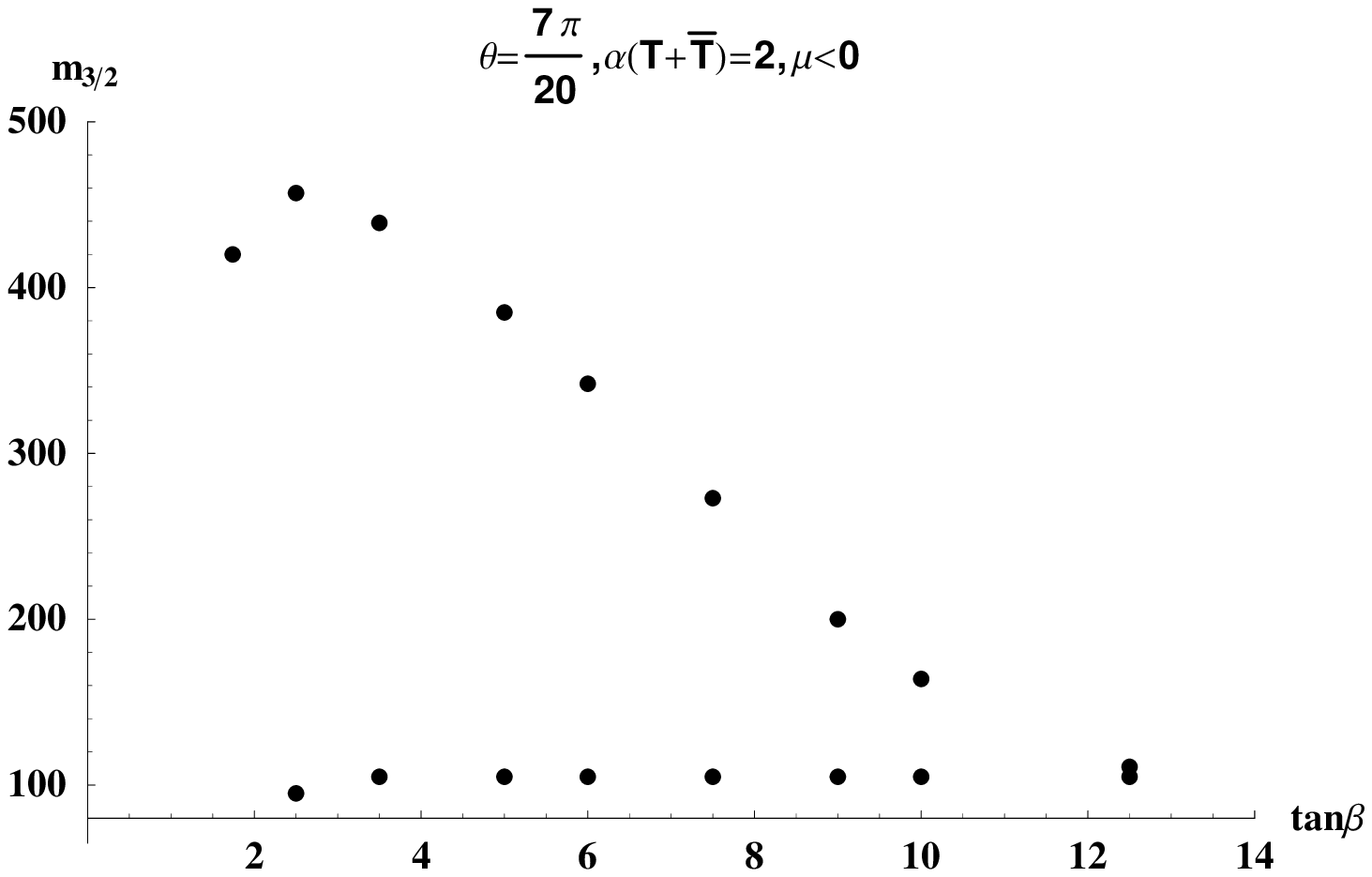}
\caption{Allowed parameter space in $m_{3/2},\tan\beta$ plane for
$\theta=\frac{7\pi}{20},\mu<0$}
\end{figure}

\begin{figure}
\epsfxsize=6in
\epsfysize=8.3in
\epsffile{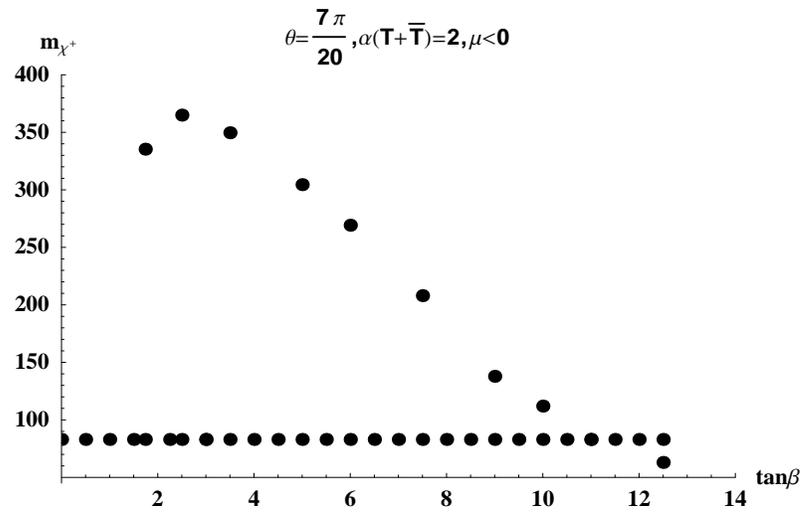}
\caption{Maximum lightest chargino mass vs $\tan\beta$}
\end{figure}

\newpage
\begin{figure}
\epsfxsize=6in
\epsfysize=8in
\epsffile{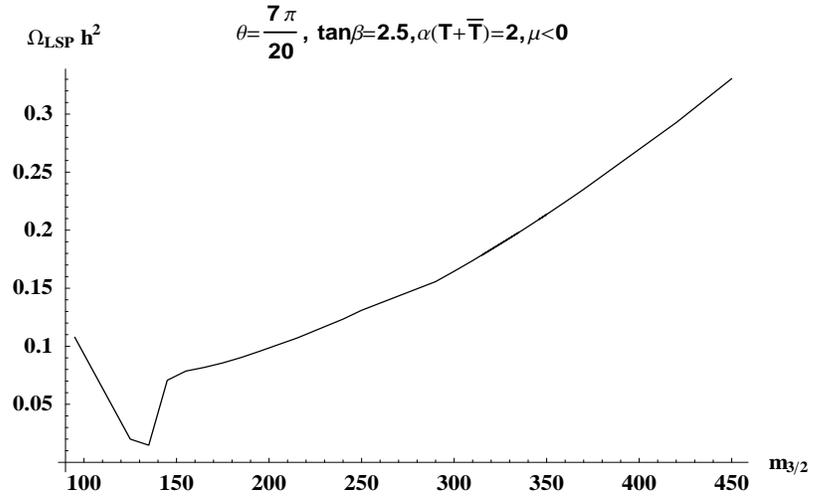}
\caption{Relic abundance versus $m_{3/2}$ for $\tan\beta=2.5,\mu<0$}
\end{figure}


\newpage
\begin{figure}
\epsfxsize=6in
\epsfysize=8.5in
\epsffile{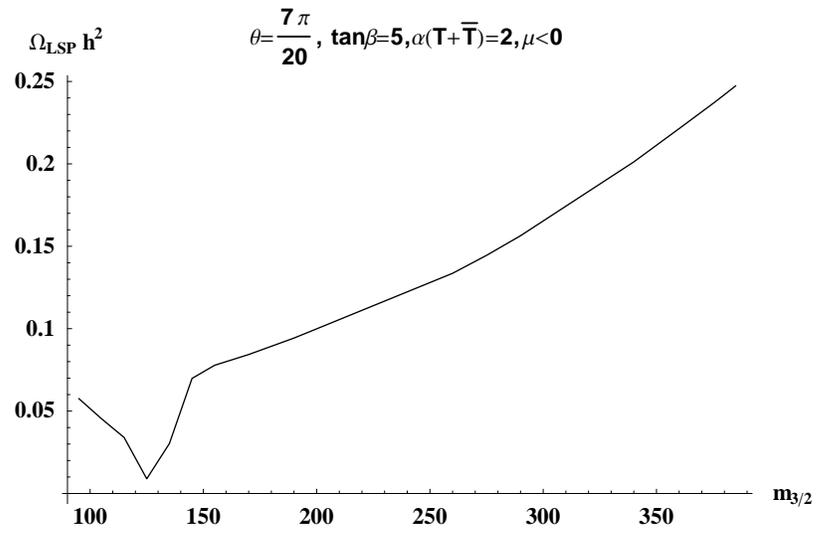}
\caption{Relic abundance of LSP vs $m_{3/2}$ for $\tan\beta=5, \mu<0$ }
\end{figure}



\newpage
\begin{figure}
\epsfxsize=6in
\epsfysize=8.5in
\epsffile{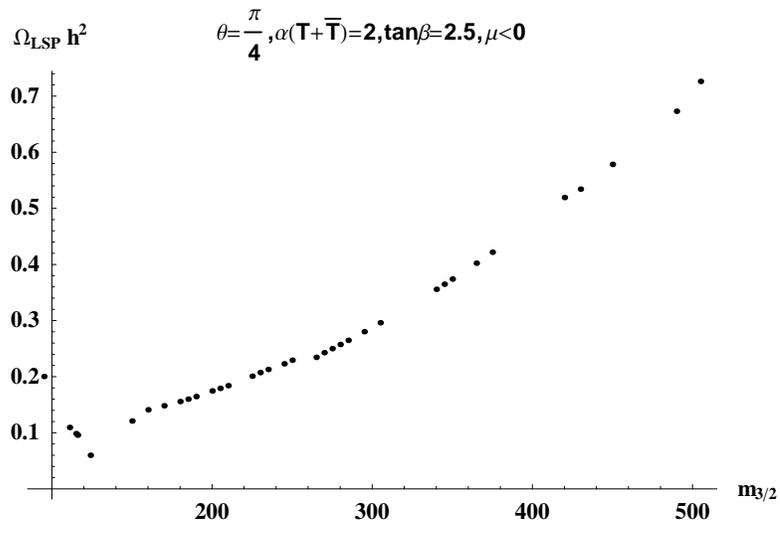}
\caption{Relic abundance of LSP in case of $\theta=\frac{\pi}{4}$ and 
$\tan\beta=2.5$}
\end{figure}





\newpage
\begin{figure}
\epsfxsize=5.8in
\epsfysize=6.7in
\epsffile{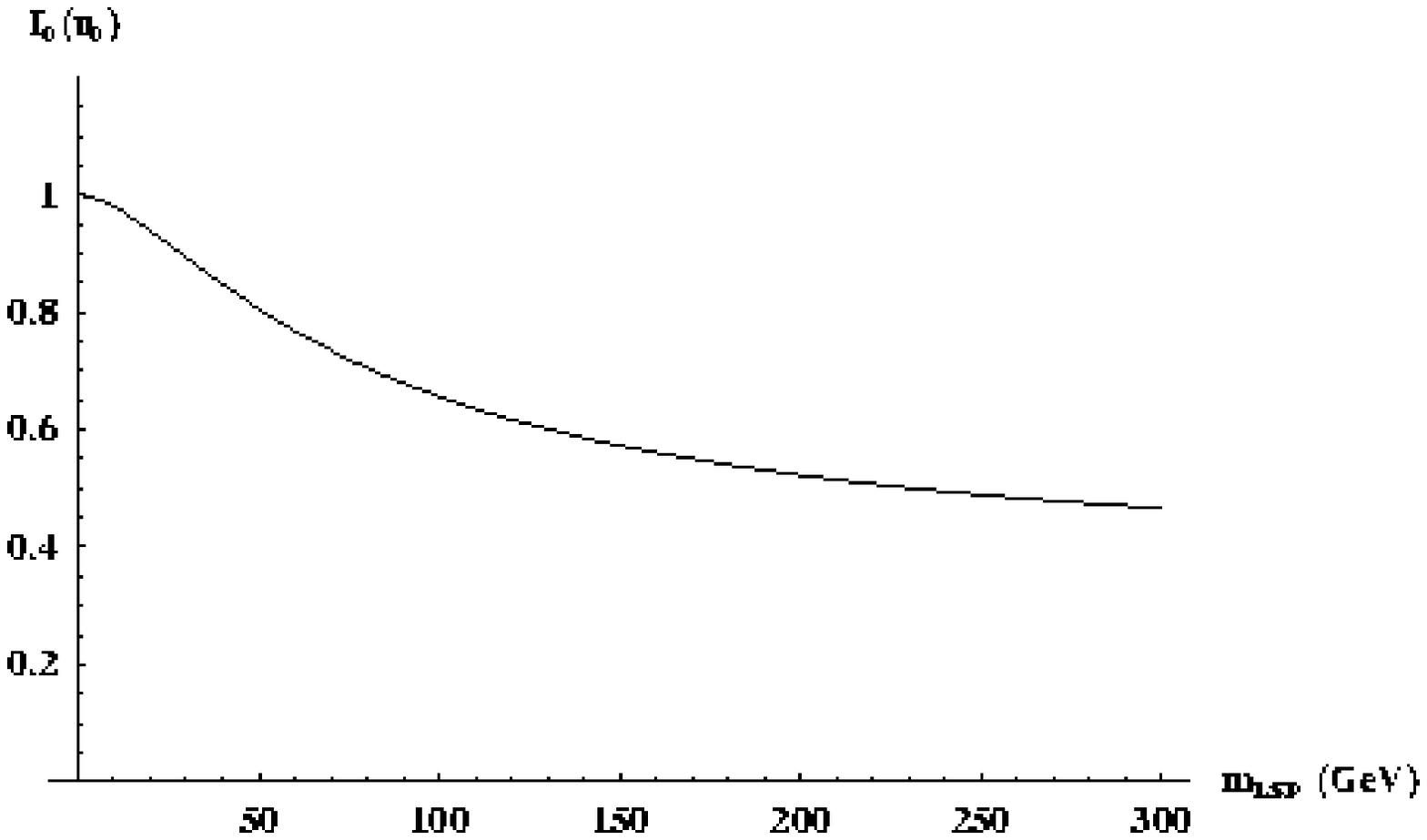}
\caption{Form factor $I_0(u_0)$ versus LSP mass for $^{72}_{32}Ge$ nuclei
see Eq.(\ref{Form})}
\end{figure}

\newpage
\begin{figure}
\epsfxsize=6.2in
\epsfysize=8.0in
\epsffile{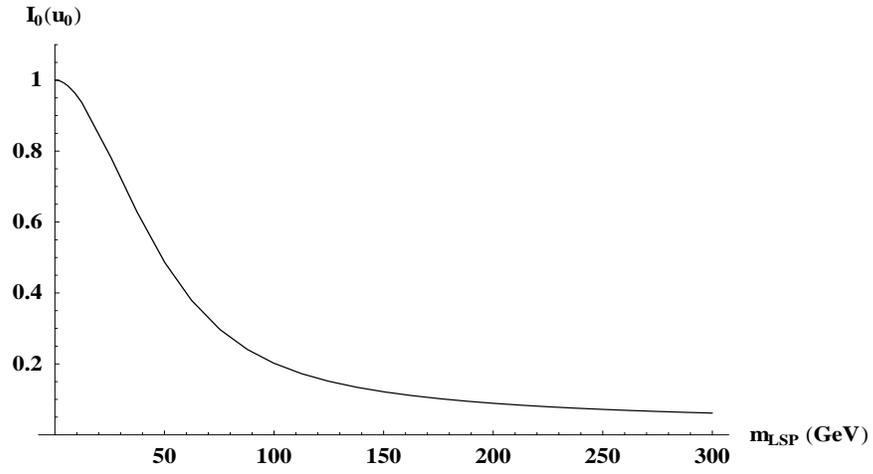}
\caption{Form factor $I_0(u_0)$ versus LSP mass for $^{208}_{82}Pb$ nuclei
see Eq.(\ref{Form})}
\end{figure}

\newpage
\begin{figure}
\epsfxsize=6.2in
\epsfysize=8.0in
\epsffile{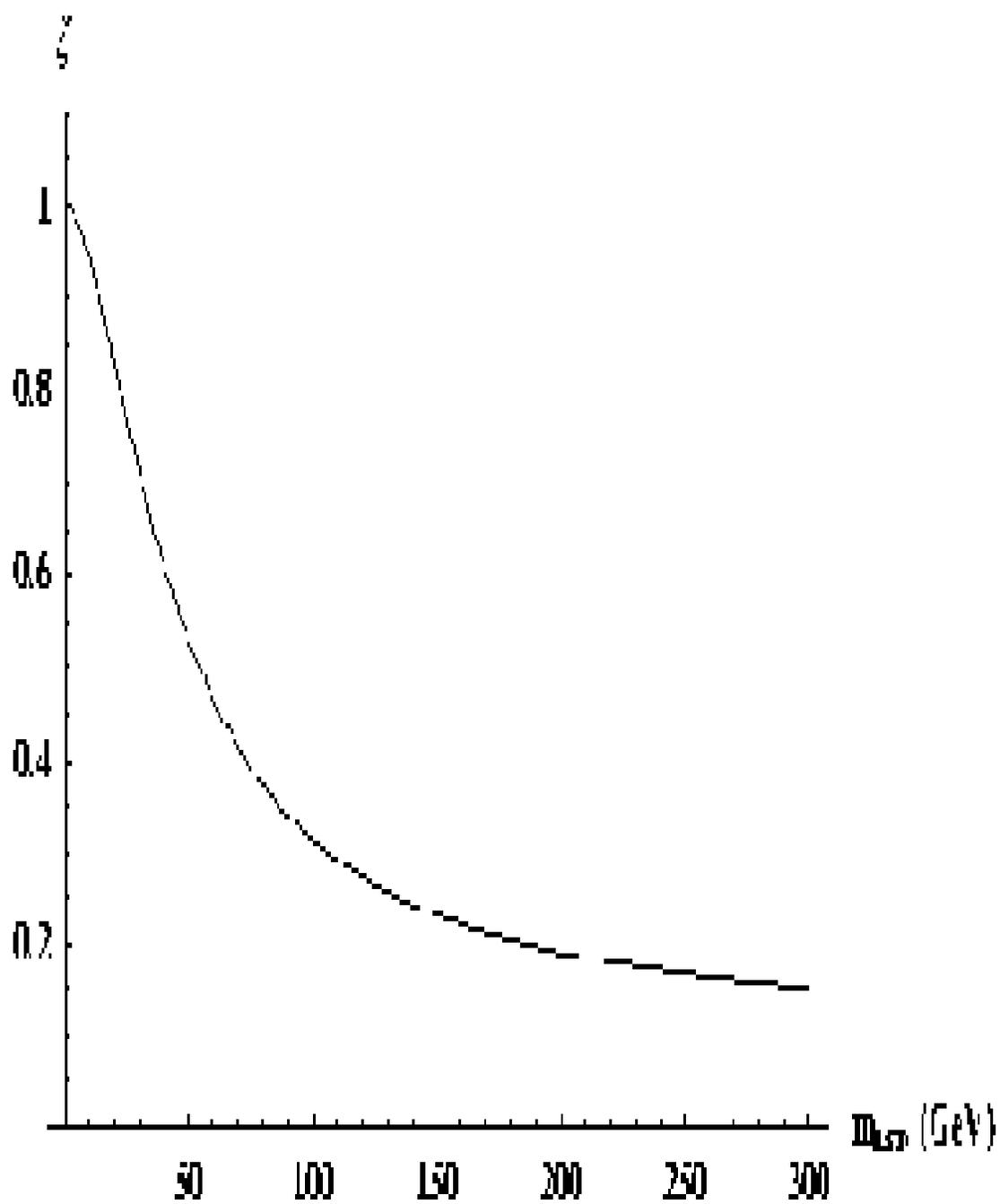}
\caption{Form factor $\zeta$ for $^{131} Xe$ nuclei}
\end{figure}

\newpage
\begin{figure}
\epsfxsize=6.2in
\epsfysize=8.0in
\epsffile{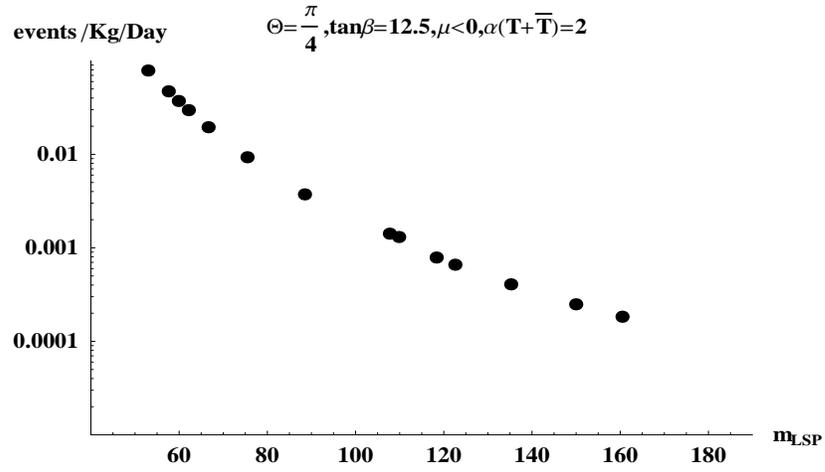}
\caption{Detection rates in the case $\theta=\frac{\pi}{4},\tan\beta=2.5$}
\end{figure}

\newpage
\begin{figure}
\epsfxsize=6.2in
\epsfysize=8.2in
\epsffile{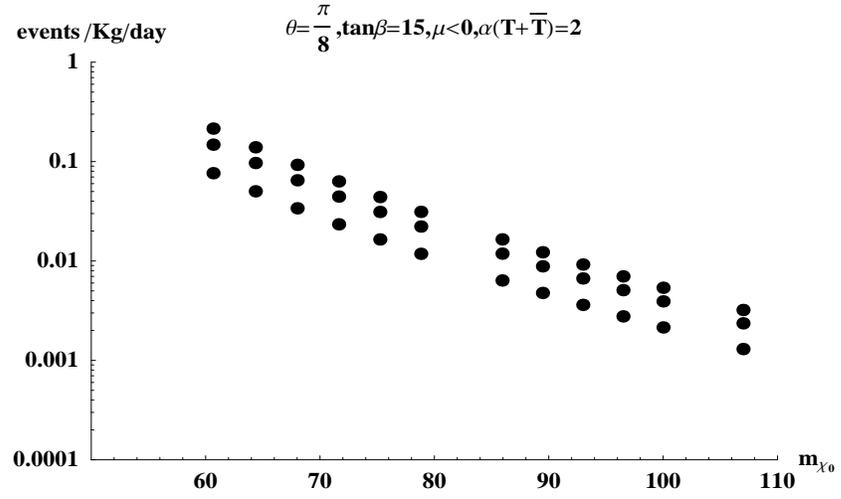}
\caption{Event rates for 3 different nuclei,$^{208}Pb,^{131}Xe,^{73}Ge$}
\end{figure}

\end{document}